# Vote Delegation in DeFi Governance

Dion Bongaerts, Thomas Lambert, Daniel Liebau, Peter Roosenboom*

*Rotterdam School of Management, Erasmus University*

March 13, 2025

**Abstract**

We investigate the drivers of vote delegation in Decentralized Autonomous Organizations (DAOs), using the Uniswap governance DAO as a laboratory. We show that parties with fewer self-owned votes and those affiliated with the controlling venture capital firm, Andreesen Horowitz (a16z), receive more vote delegations. These patterns suggest that while the Uniswap ecosystem values decentralization, a16z may engage in window-dressing around it. Moreover, we find that an active and successful track record in submitting improvement proposals, especially in the final stage, leads to more vote delegations, indicating that delegation in DAOs is at least partly reputation- or merit-based. Combined, our findings provide new insights into how governance and decentralization operate in DeFi.

* All authors are with the Rotterdam School of Management, Erasmus University, Burgemeester Oudlaan 50, 3062 PA Rotterdam, the Netherlands, emails: dbongaerts@rsm.nl; t.lambert@rsm.nl; liebau@rsm.nl; proosenboom@rsm.nl. We thank Shiwei Ye for excellent research assistance and the team at butter.money for their support in extracting data from the Ethereum blockchain. We also acknowledge the FinTech and Digital Finance Chair at Paris Dauphine University for financial support. We thank seminar participants at City University of Macau for their valuable comments. The usual disclaimers apply.

## 1. Introduction

Decentralized Finance (DeFi) represents a new financial paradigm that leverages distributed ledger technologies to provide services such as lending, investing, and trading without reliance on traditional centralized intermediaries (Auer et al., 2024). Among the most significant innovations in DeFi are decentralized exchanges (DEXs), which replace central limit order books with automated market makers (AMMs) that determine prices algorithmically (Capponi and Jia, 2021; Harvey et al., 2024). Uniswap, widely regarded as the most prominent decentralized exchange, has been at the forefront of these developments (Lehar and Parlour, 2025).

DeFi has garnered substantial attention and investment, with total value locked—the capital deployed on DeFi platforms—surging from under USD 1 billion in May 2020 to just above USD 145 billion in April 2022, according to analytics platform DeFiLlama. At the same time, DeFi has sparked debates over governance and control. A key regulatory concern is whether these platforms are "sufficiently decentralized," a designation that can exempt them from certain legal frameworks, including the EU's Markets in Crypto-Assets (MiCA) regulation. However, the criteria for determining "sufficient decentralization" remain ambiguous, raising fundamental questions about governance structures in DeFi ecosystems.[2]

Unlike traditional financial intermediaries, which are organized as corporations, most DeFi platforms rely on Decentralized Autonomous Organizations (DAOs) for governance. DAOs use smart contracts and blockchain technology to facilitate collective decision-making (Hsieh et al., 2018; Hassan and De Filippi, 2021; Ding et al., 2023; Han et al., 2025). They replace managers and directors with majority voting and coalition building (Werbach et al., 2024). This governance structure enables coordination without a central authority and may reduce traditional agency costs, as token holders ("shareholders") can directly manage their organization by voting on proposals. Successful DAOs have active voter communities (Laturnus, 2023).

[2] Source: https://blog.merklescience.com/general/is-defi-truly-exempt-from-mica-regulations (last accessed: December 10, 2024).

Many leading DeFi platforms on the Ethereum network implement governance using the Governor Bravo smart contract. A key feature of this smart contract is decentralized decision-making through voting, similar to how shareholders influence corporate decisions through voting or board elections. To enhance participation and mitigate free-riding, Governor Bravo, in combination with the respective Governance token smart contract of the DAO, allows token holders to delegate their voting power to others. In some of the most important DAOs in DeFi, token holders who choose not to vote directly can delegate their votes to representatives, known as delegates, who vote on their behalf (Nazirkhanova et al., 2025). These delegates play a crucial role in governance, as they represent token holders', or delegators', interests.

Vote delegation is a critical yet understudied aspect of DAO governance, which we explore in this paper. Specifically, we investigate the factors driving vote delegation decisions, using the Uniswap DAO as a laboratory. Governance power in Uniswap is tied to UNI tokens, which holders can use to vote on proposals. Alternatively, they can delegate their voting rights to other addresses. While Uniswap is the largest decentralized exchange, the Governor Bravo contract—on which its governance relies—is also used by several other major DeFi platforms, including Aave, Compound Finance, Sushiswap, and Balancer. The widespread adoption of Governor Bravo and the way to handle delegations in the respective token contract enhances the generalizability of our findings.

We examine vote delegation patterns in DAOs by testing three hypotheses. First, we hypothesize that UNI token holders are more likely to delegate voting rights to parties with a significant ownership stake (i.e., wallets with large UNI token holdings). Since ownership stakes align decision-makers' incentives with stakeholder interests, we expect token holders to prioritize delegates with substantial holdings—that is, skin in the game. When parties have little or no ownership stake, their interests may diverge from those of other stakeholders (Cremers et al., 2009). Thus, token holders are likely to delegate their votes to delegates whose substantial stake signals alignment with collective interests.

Second, we hypothesize that if decentralization is a key priority for the ecosystem, large stakeholders will delegate voting rights to independent parties with the freedom to vote autonomously. This hypothesis builds on organizational theories, which emphasize the critical monitoring role of independent decision agents (Fama and Jensen, 1983). Independent parties enhance oversight by providing checks and balances that mitigate agency problems. Delegating voting rights to independent parties introduces external perspectives and reduces the risk of decision-making being dominated by narrow interests.

Third, we hypothesize that delegation of voting rights is influenced by demonstrated skill and commitment, leading to a reputation- and merit-based delegation model. Optimal delegation favours agents with strong reputations for relevant expertise (Aghion and Tirole, 1997). In DAOs, participants with a track record of technical expertise, community leadership, and ecosystem contributions are more likely to receive delegated votes. This means that for delegates, successfully proposing and having those proposals passed builds reputation (a reputational premium *à la* Guerieri and Kondor, 2012), which in turn attracts additional voting power from others.

We test our hypotheses using novel data on vote delegations. Our sample consists of wallets that, at any point between September 2020 and April 2022, were among the top 5 voters for a given voting event (we coin this group the "Top5" voters). We focus on these wallets because the literature shows (and we confirm) that voting in DAOs tends to be highly concentrated (e.g., Lustenberger et al., 2024). We dedicate substantial effort in collecting data on the characteristics and voting history of these wallets from on- and off-chain data.[3] Our final sample includes 47,808 daily observations of vote delegations (to oneself and others), as well as voting activity across all stages of the governance process for each of the 48 proposals submitted during our sample period.

Our regression analysis yields three key results. First, we do not find support for our first hypothesis that wallets with larger UNI token holdings should attract more delegated votes than those with smaller holdings. We find a negative association between self-delegated voting power and the votes received from others—a 1% increase in self-delegation is associated with a 1.58% to 1.63% decrease in votes from others. This result challenges traditional governance theories, which emphasize skin-in-the-game effects, and instead suggests that the ecosystem prioritizes decentralization.

[3] We manually collected these data from the Messari platform and using scripts from the Dune platform, which gives access to the Ethereum blockchain directly.

Second, we show the central role played by early investor Andreessen Horowitz (a16z) in Uniswap governance, not only through direct voting but also via a network of aligned delegates. Being an a16z affiliate significantly increases the votes received from others, by factors between 2,689 and 3,547. This coordination is also evident in voting patterns: in on-chain votes where Andreessen Horowitz participated, affiliated delegates voted against them only once out of eleven voting events. Our analysis reveals a governance structure that preserves the appearance of decentralization while enabling tight control within the ecosystem. Combined with the first result on the appreciation of decentralization, this finding suggests window dressing.

Last, we find a positive association between (successful) proposal-making activity and voting power based on delegated votes from others. Proposers receive around 59 times more delegated votes. The effect is mainly driven by successful proposers and is also stronger in the final stage of the governance process—the on-chain vote that directly implements the proposal on the blockchain. This result extends our understanding of governance participation by highlighting how a proven track record in proposal-making and success significantly influence delegation decisions, supporting the idea that merit and reputation play a crucial role in attracting additional delegators.

Our study draws on the emerging literature on DeFi (Chen and Bellavitis, 2020; Schär, 2021; John et al., 2023). Within this literature, several papers examine the true extent of decentralization in DeFi.[4] Anker-Sørensen and Zetsche (2021) introduce the concept of "Fake-DeFi" that describes DeFi platforms appearing decentralized but in reality being centrally controlled. In the specific context of DAO governance, a few papers provide empirical descriptions of the distribution of votes. Fritsch et al. (2024) analyse voting power in three DAOs, including Uniswap, and describe the structure of delegation networks by distinguishing "single holder"—those who receive at least 50% of their voting power from a single token holder—from the remaining "community" delegates. They find that large and powerful delegates typically decide in the same way as the larger community. Appel and Grennan (2023) review proposals across 151 DAOs, finding that only a small number of entities exert control over most decisions. Laturnus (2023) also documents the concentration of voting in DAO proposals, while Sun et al. (2024) present evidence of centralization in tokenized voting within MakerDAO polls. Bakos and Halaburda (2022) develop a model of democratic DAO governance and find that making governance tokens *tradable* can lead to the centralization of voting power, as large token holders can obtain greater influence through trading.[5] Bellavitis and Momtaz (2024) examine how specific DAO characteristics, such as ownership concentration, affect the efficacy of on-chain versus off-chain voting. Feichtinger et al. (2023) and Barbereau (2023) document relatively low governance participation by regular token holders and delegates. We build on and extend these studies by showing how delegation- and token-holding decisions impact delegates' ability to accumulate voting power from others. To our knowledge, we are the first to distinguish between self-delegation versus delegation from others, which is crucial to further our understanding of actual delegated decision-making in a DAO context. Our findings also highlight the role of merit and reputation in attracting additional delegators, echoing research on their roles in funding dynamics in crowdfunding and token offerings (e.g., Kim and Viswanathan, 2019; Lee et al., 2022; Belleflamme et al., 2025).

[4] For an excellent discussion of decentralized versus distributed organizations, see Vergne (2020).

Our work adds more broadly to the literature studying DEXs. While several other papers study the impact of DEXs on liquidity and market quality (e.g., Park, 2023; Lehar and Parlour, 2025), we study the governance side of DEXs, which are typically organized through DAOs.

Since Uniswap operates as a DAO, with no managers or directors, our work also contributes to the literature on corporate governance and shareholder democracy.[6] Gantchev and Giannetti (2021) explore how the voting process filters out ill-informed shareholder proposals, finding that a significant proportion of proposals come from a few active sponsors and that these proposals are less likely to gain majority support. However, they can still pass if supported by a majority of arguably uninformed shareholders.[7] Although we also observe that in Uniswap most proposals are submitted by a limited number of sponsors, we do find evidence that support by a majority of uninformed token holders is not required for passing and that rather, support from the Top 5 voters is typically sufficient. In contrast to studies that focus on shareholder proposals that never reach the voting stage (see, e.g., Soltes et al., 2017; Matsusaka et al., 2019), our analysis using Uniswap data further shows that a significant 27% of proposals submitted make it to the final stage. By focusing on (delegated) voting, our study also belongs to the theoretical literature on shareholder voting (see Levit et al., 2024, among others). Interestingly, Gersbach et al. (2022) find that vote delegation tends to produce more favourable outcomes with higher probability than conventional voting in costly voting environments, especially when malicious voters are present and preferences are private. Their findings have direct implications for DAO voting schemes, suggesting that delegation can have negative outcomes if the number of misbehaving agents exceeds a critical threshold.

[5] Relatedly, Han et al. (2023) theoretically show potential conflicts of interest between large and many small participants in DAO governance, while Aoyagi and Ito (2022) focus on the competition among DAOs.
[6] Studying DAO governance is indeed relevant for the governance of traditional corporations. For example, every participant in a DAO has a financial stake and shares the risk of default, it is therefore expected that all members align around a common organizational goal and are motivated to enhance the overall value of the DEX. By linking control to ownership, DAO members can collectively manage the organization, prioritize shared interests, and reduce the traditional conflicts between owners and managers (Becht et al., 2003).
[7] Several other papers investigate the effects of mutual funds' attributes on voting behaviour (see, e.g., Cvijanović et al., 2016; Malenko and Shen, 2016).

## 2. Uniswap and its Governance

Uniswap is a DEX launched in November 2018 that enables users to swap tokens without intermediaries by directly settling transactions on the Ethereum blockchain. In early 2019, its founder, Hayden Adams, incorporated a Delaware-based company, Uniswap LLC (later renamed Universal Navigation Inc.), and secured USD 1.8 million in funding from the cryptocurrency hedge fund Paradigm. By June 2020, Uniswap raised an additional USD 11 million from Andreessen Horowitz (a16z), granting the venture capital firm a significant allocation of governance tokens (UNI). In September 2020, Uniswap formally launched its governance process alongside the UNI token (Kim, 2021). The platform's popularity surged, reaching over USD 9 billion in total value locked (TVL) in its smart contracts by early 2021.[8] Today, Uniswap remains the most prominent DEX within the Ethereum ecosystem and beyond.[9]

This study examines the governance structure of the Uniswap DAO. The DAO issues UNI tokens, which grant holders voting rights and plays a central role in governing the Uniswap protocol. Governance decisions fall into two broad categories. The first involves technical and economic decisions, such as protocol configuration changes—for example, enabling token holders to receive a share of the protocol's fee income. The second category pertains to financial decisions, where DAO members propose initiatives requiring funding from the protocol's treasury, which consists of mostly UNI tokens, accessible via the governance process. Members then cast votes to determine which initiatives receive funding. In this context, the Uniswap Grants Program (UGP) was introduced in December 2020 to support projects that contribute to the Uniswap ecosystem.

[8] TVL is a metric that refers to the total amount, denominated in USD, that is locked or staked in a DeFi protocol. It is often used to assess the importance of a DeFi protocol relative to others or as a gauge to analyze the overall DeFi ecosystem.
[9] Uniswap is now also available on various other blockchains like Arbitrum, Base and others, https://support.uniswap.org/hc/en-us/articles/14569415293325-Networks-on-Uniswap (last accessed: December 10, 2024) provides the full list.

The UNI token is essential for submitting and voting on governance proposals. However, merely holding UNI tokens is insufficient to participate; token holders must first delegate their voting rights. Delegation can be either to oneself (self-delegation) or to another wallet (delegation-to-others) via a transaction on the Ethereum blockchain. A wallet owner must delegate their entire UNI balance, as fractional delegation is not possible.[10] Any Ethereum wallet address can serve as a delegate. Once voting rights are delegated, the delegate wallet can propose and vote on governance initiatives. Delegation can be reversed, but the token holder must sign a blockchain transaction—incurring a gas fee—to reassign voting rights. Notably, delegation does not automatically update when UNI tokens are sold or transferred. For example, if wallet A, holding 100 UNI tokens, delegates its voting rights to wallet B, wallet B retains voting power even if wallet A transfers its tokens to wallet C. The voting power remains with wallet B until wallet C actively claims it through self-delegation or assigns it to another delegate. This phenomenon is further explained in the Appendix A.

Uniswap core team members—who have often received substantial allocations of UNI tokens as compensation for their contributions—have publicly committed to abstaining from the governance process, including voting and submitting proposals. While verifying adherence to this self-imposed restriction on a case-by-case basis may be challenging, the alignment of incentives and the high stakes involved serve as deterrents. Any evidence to the contrary uncovered by the broader token-holder community could significantly harm Uniswap's reputation.

The governance process is designed to empower the community of UNI token holders and delegates to make decisions regarding the development of the Uniswap protocol and its broader ecosystem. Community members can propose initiatives, which are then subject to a voting process that determines their acceptance or rejection. Governance follows a four-step procedure, three of which involve voting.[11] Table 1 summarizes the stages of the governance process.

[10] Of course, multiple wallets can be created to work around this constraint.

The first step involves proposers sharing their ideas on Uniswap's online discussion forum. Specifically, they must submit a post in the "Proposal Discussion" section, outlining their proposal at a high level and inviting initial feedback. To maximize community engagement, Uniswap encourages proposers to contact Discord administrators to organize a community call. In addition, proposers often leverage social media platforms such as X (formerly Twitter) to increase visibility and solicit input and feedback.

The second step requires the proposer to create a forum post in the "Temperature Check" section, incorporating feedback from the initial discussion. Simultaneously, the proposer must publish a corresponding "Temperature Check" vote on the Snapshot platform, marking the first formal voting event. Snapshot is a decentralized voting system that eliminates the need for voters to pay Ethereum gas fees, which can be volatile and costly, thereby encouraging broader participation. The system records voting results on the InterPlanetary File System (IPFS) and immutably stores only the final results—not individual votes—on the Ethereum blockchain, ensuring verifiability and reducing the risk of disputes. To submit a proposal on Snapshot, proposers must connect using an Ethereum wallet, optionally utilizing the Ethereum Name Service (ENS) for identity disclosure. However, anonymity remains an option. The temperature-check voting period lasts at least two days, requiring a minimum of 25,000 affirmative votes to advance to the next stage. If the proposal fails to meet this threshold, it is rejected, and no changes are implemented.

In the third step, the proposer updates the Temperature Check forum post to reflect community feedback and initiates a second Snapshot vote, known as the "Consensus Check." This marks the second voting event and lasts for a minimum of five days. To advance to the final stage, the Consensus Check must receive at least 50,000 affirmative voting tokens. If this threshold is met, the proposer can proceed to the final step of the Uniswap governance process: the on-chain vote. If the proposal fails to secure sufficient support at this stage, it is rejected, and no changes are made to the system.

[11] In December 2022 (after our sample period) the Uniswap community put forward a proposal to simplify the governance proposal process. Source: https://gov.uniswap.org/t/proposed-simplifications-to-the-uniswap-governance-process/15906/3 (last accessed: May 5, 2024).

The fourth and final step in the governance process is the on-chain vote. The Governor Bravo smart contract facilitates the management of this part of the process in conjunction with the UNI token smart contract that keeps track of vote delegations.[12] To submit a proposal on-chain, the proposer must control at least 2.5 million voting tokens and initiate the vote by signing a transaction on the Ethereum network, which incurs a gas fee. The voting period lasts a minimum of seven days.[13] For a proposal to pass, it must secure a majority of affirmative votes while meeting a quorum of at least 4% of the total voting power—equivalent to 40 million votes during our period of analysis. Furthermore, before an on-chain vote occurs, the proposer must submit audited source code that implements the proposed change. If approved, the code is automatically deployed after a two-day delay period, known as a timelock, which is enforced via a smart contract. This delay serves as a security mechanism, allowing the Uniswap community to review the decision and, if necessary, propose and approve a revised version to override a potentially malicious or harmful proposal. If the on-chain vote fails to meet the required threshold, the proposal is rejected, and no changes are made to the system.

## 3. Data

### 3.1. Sample and voting events

We use several sources to assemble a panel data set at the wallet and day levels. Our analysis spans from 17 September 2020, when the Uniswap Governance process was launched, to 15 April 2022.

We hand-collect information on Uniswap governance voting events from the Messari platform using its "Governor" functionality (different from the Governor Bravo smart contract mentioned earlier). We exclude 4 events from the 52 Uniswap voting events in our sample period due to data incompleteness, resulting in 48 events. These comprise 21 temperature checks, 14 consensus checks, and 13 on-chain votes.[14] Of these 48 events, 38 (or 79%) succeeded, while ten failed to achieve the required support or were cancelled by their proposers. Bellavitis et al. (2023) observe a similar approval rate of 87% when documenting basic facts of a cross-section of DAOs.

[12] Governor Bravo is the smart contract that manages proposal creation, voting, and execution. It enables DAOs to configure key parameters such as voting period duration, timelock duration, proposal threshold, and quorum, working in tandem with the UNI token smart contract which tracks vote delegations.

[13] In January 2024 (after our sample period), the Uniswap community approved a proposal to change the threshold to propose on-chain votes from 2.5 million to 1 million. Source: https://app.uniswap.org/vote/2/55 (last accessed: May 5, 2024).

The Messari platform provides information on the five most significant voters by voting power for each voting event. Our sample identifies 84 unique wallets that appear in this "Top5" list for at least one voting event. For each of those wallets, we then retrieve daily data on delegation amounts (self-delegated and from others) from the Ethereum blockchain, leading to a total of 48,384 observations in our sample. In our analysis, we run all regressions excluding the a16z proprietary wallet, which controls up to 15 million votes owned by the venture capital firm and receives only minimal delegation from others during our analysis period.[15] Excluding this wallet reduces the total number of observations to a maximum of 47,808 and prevents our results from being driven by this outlier. However, our results remain robust to the inclusion of the a16z wallet.

The distribution of voting power among the "Top5" voters reveals a high degree of concentration. In each voting event, the voter with the highest share of votes cast is referred to as the Top1 voter. In our data set, Top1 voters, on average, control approximately 7 million voting rights, representing 46% of all votes cast across voting events. Similarly, we define Top2, Top3, Top4, and Top5 voters based on their respective vote shares in each event. As illustrated in Figure 1, the proportion of votes cast declines sharply beyond Top1, with Top5 voters commanding only 4% of votes on average.

Figure 1 further shows that, on average, the Top1 and Top2 voters collectively control 65% of the votes cast, granting them substantial influence over voting outcomes. This finding aligns with Grennan and Appel (2023) who show that in many DAOs, as few as two dominant voters can effectively determine the results of governance decisions.

[14] The proposal subject to these events address critical matters; for instance, deploying the Uniswap AMM to blockchains like Arbitrum (generating over USD 600 million in trading volume by 2025) or allocating USD 20 million to the DeFi Education Fund to represent Uniswap's interests at Capitol Hill.

[15] That the Andreessen wallet receives only minimal delegations from others is in line with another result we document: Declaring one's identity through the use of an ENS name does not enter our specifications significantly.

Figure 2 presents the Herfindahl-Hirschman Index (HHI) of voting power concentration over time, revealing a gradual decline throughout our sample period. However, despite this downward trend, the concentration of voting power remains sufficiently high to raise fundamental questions about whether DAOs genuinely achieve their decentralization objectives.

Each of the voting events collected from Messari also contains information on the respective proposers. We identify 25 unique proposers, 17 of which are part of the 84 Top5 voters (un-tabulated statistics).

### 3.2. Variable descriptions

We employ a wide range of variables, capturing voting outcomes, delegate characteristics, market conditions, voting-period metrics, and proposer-related information. Table 2 provides a detailed definition of all variables.

We are primarily interested in understanding what drives votes to be delegated to a specific delegate. To focus on delegation decisions that are material for economic outcomes, we restrict ourselves to delegations towards wallets that ever show up as Top5 voters. Our main outcome variable is the log of the number of votes delegated from others to each delegate. As delegate characteristics, we consider delegates' own voting power (log of votes delegated to oneself), dummy variables of delegate identity; that is, whether the wallet has an ENS name associated with it (Nickname), and whether it has the early investor affiliation (a16z Affiliate). To identify a16z affiliates we refer to the investor webpage and then look up wallets based on matching ENS names.[16]

We also construct market-related variables, for which we obtain data from CoinGecko starting from September 19, 2020, primarily to serve as market-wide controls. CoinGecko aggregates market data from centralized exchanges (e.g., Coinbase, Binance) and decentralized exchanges (e.g., Uniswap), providing coverage of trading activity across venues. The market-wide control variables we collect also include the log of the UNI token market capitalization, trading volume, the Fear & Greed index, and the ETH price. We also collect the average transaction fee variable because it could shed light onto the urgency of vote control for token holders since fees in the Ethereum ecosystem are known to be highly volatile and both un-delegations and re-delegations incur gas fee payments.

---

[16] The link to the relevant a16z web-page is: https://a16z.com/open-sourcing-our-token-delegate-program. The affiliated delegates are: 0x2b1ad6184a6b0fac06bd225ed37c2abc04415ff4 (Andreessen Horowitz), 0x5d8908afee1df9f7f0830105f8be828f97ce9e68 (Argent),0x458ceec48586a85fcfeb4a179706656ee321730e (Blockchain at Berkeley), 0xdc1f98682f4f8a5c6d54f345f448437b83f5e432 (Blockchain at Columbia), 0x13bdae8c5f0fc4023d1f0e6a4ad70196f59138548 (Blockchain at Michigan), 0x47c125dee6898b6cb2379bcbafc823ff3f614770 (Blockchain at UCLA), 0x7e4a8391c728fed9069b2962699ab416628b19fa (Dahrma), 0x6626593c237f530d15ae9980a95ef938ac15c35c, proxy for 0x683a4f9915d6216f73d6df50151725036bd26c02 (Gauntlet), 0x9b68c14e936104e9a7a24c712beecdc220002984 (Getty Hill), 0x61c8d4e4be6477bb49791540ff297ef30eaa01c2 (Harvard Law BFI), 0x070341aa5ed571f0fb2c4a5641409b1a46b4961b (Penn Blockchain), and 0xeff506a32b55d5c19847c1a4f8510c00280c27e5 (Stanford Blockchain Club).

We include a set of variables related to the voting period to capture the effect of governance activity timing on delegation decisions. Specifically, we incorporate dummy variables for active proposal periods at different stages (temperature check, consensus check, and on-chain vote), as well as a dummy variable indicating whether any proposal period is active.

We further rely on proposer-related variables. We measure reputation building by counting the number of times the 84 Top 5 wallets in our data set propose at the different stages in the governance process. We also include a set of dummy variables capturing wallet proposal participation across various voting event types, such as whether a wallet acted as a proposer at any stage—specifically for temperature checks, consensus checks, or on-chain votes—and whether these proposals were successful. These variables provide insights into whether having proposed and having successfully proposed influences subsequent vote delegation to the delegate. In addition, we track cumulative proposal activity through counts of proposals a delegate has made at each governance stage (temperature check, consensus check, and on-chain vote) and the total number of proposals across all stages.

### 3.3. Descriptive statistics

Table 3 presents descriptive statistics for the variables in our analysis. Our outcome variable is the log of votes delegated from others, measured at the wallet-day level. The log-transformed variable has a mean of 4.666 (or 1,371,854.09 actual votes) and standard deviation of 6.751 (2,885,099.54 votes), with a median of zero, representing observations with no delegated votes. While on some days some wallets show significant delegation activity, the typical daily state for most wallets is to control no delegated votes.[17] Meanwhile, the maximum equals 16.578 (or 15,845,481.7 in actual votes). These figures suggest that a handful of delegates control substantially more voting power than the rest (as also shown by Figures 1 and 2) and reflects the fact that several of the 84 wallets early on had no external delegations.

The delegates' own voting power, measured as the log of votes delegated to oneself, has a mean of 1.478 (443.67 actual votes) and standard deviation of 2.813 (or 2,676.810 actual votes), which is much smaller than voting power received from delegation. Furthermore, 50.6% of delegates in our data set have a nickname, and 18.1% are categorized as a16z affiliates.

For voting-period variables, we observe that temperature check periods are active for 5.7% of observations, consensus check periods on 8.3%, and on-chain vote periods also on 8.3% of observations, with at least one type of proposal period being active for 16.7% of the observations in the data set.

We observe that 8.8% of the 47,808 wallet-day observations are associated with Top5 wallets launching one or more proposals. The percentages fall from the initial stages of the governance process: 5.5% of wallet-day observations are associated to temperature checks, 3.5% to consensus checks, and 3.6% to on-chain votes. Percentages are even smaller for wallet-day observations associated to successful proposals — 5.5% of all wallet-day observations are associated to successful proposals. Again, the percentages diminish through the governance stages, 3.4% are associated to success at temperature check-level, 2.3% at consensus-check-level, and 2.4% at on-chain vote-level. Finally, we count the number of wallets-day observations that are associated with proposals for each wallet at the various levels of the governance process. Only a small group of wallets propose at all, and even fewer wallets do so multiple times which explains the average of 0.157 proposals per wallet-day observations for the temperature check stage, 0.108 for the consensus check stage, and 0.072 for the on-chain vote stage.[18]

[17] The Figure in Appendix B provides the density plot of the outcome variable, Votes Delegated from Others.
[18] Actual success rates per governance stage are 71.4% (temperature check), 85.7% (consensus check), and 76.9% (on-chain vote).

## 4. Empirical Analysis

To better understand delegation dynamics, we compare delegated voting power from others with self-delegation over time. Figure 3 shows that between September 2020 and April 2022, self-delegated voting power averaged only 0.04 million, whereas delegated voting power from others averaged 128.09 million.

We examine the factors influencing vote delegation from other UNI token holders using a Tobit model to account for the left censoring of our outcome variable, typically at zero.[19] Our main specification is:

$$Votes\ Delegated\ from\ Others_{i,t+1} = \alpha + \beta DelegateCharacteristics_{i,t} + \gamma X_{i,t} + \delta_j + \varepsilon_{i,t}, \qquad (1)$$

where $Votes\ Delegated\ from\ Others_{i,t}$ is the log of daily number of votes delegated by other wallets to delegate $i$ on day $t$ plus one. $\alpha$ is a constant term. $DelegateCharacteristics_{i,t}$ contains the variables capturing delegate characteristics, namely the log the votes delegated to oneself, the Nickname dummy variable, and the a16z affiliate dummy variable. The vector of explanatory variables, $X_{i,t}$, always includes the market-related variables: the log the UNI token market capitalization, the log of one plus the UNI total trading volume, the Fear and Greed index, the log of the ETH price, and the log of the average transaction fee. In some specifications, it also includes the voting-period dummy variables for each proposal period (temperature check, consensus check, on-chain vote) and for the active voting periods. In some other specifications, it also further adds the proposer-related variables, including the voter's track record in being a proposer as a proxy for reputation. All specifications also include $\delta_j$ denoting the week that day $t$ is part of. $\varepsilon_{i,t}$ is the error term. We cluster standard errors at the delegate level to account for within-delegate correlation and make them robust to heteroscedasticity.

We test three hypotheses about governance participation and influence in Uniswap. First, we hypothesize that, in line with a "skin in the game" dynamic, higher self-delegated voting power leads to greater voting power attracted from others. Second, we expect major initial investors, particularly Andreessen Horowitz, to actually maintain distance from day-to-day governance processes in support of protocol decentralization and delegate to wallets that can freely vote. Third, we hypothesize that delegation decisions are merit- or reputation-based. Specifically, for delegates, successfully proposing and having those proposals passed builds reputation, which in turn attracts additional voting power from others.

[19] The estimation results are qualitatively similar if an OLS model is used.

Table 4 presents our main regression results examining the factors driving vote delegation. We first find across all specifications that a 1% increase in votes delegated to oneself is associated with a 1.58-1.63% decrease in votes (or approximately a decrease of 22,000 votes) received from others, with the effect being statistically significant at the 1% level. This result is not in line with our first hypothesis, relating to having "skin in the game", where higher self-delegated voting power leads to greater voting power attracted from others. One possible explanation for this finding is that the community of delegators values decentralization more than skin in the game and therefore prefers to delegate to wallets that do not hold significant amounts of UNI tokens themselves.

While having a nickname may signal transparency to potential delegators, we find no statistical evidence that this characteristic significantly influences the amount of voting power received from others. However, being an a16z affiliate has a substantial effect—an a16z affiliate receives about 2,689 and 3,548 times the votes compared to a non-affiliated delegate, with an effect being statically significant at the 1% level across all specifications.[20] This delegation pattern raises questions about the actual decentralization of governance in the Uniswap DAO.

The strong influence of a16z and its affiliated members is consistent with a window-dressing strategy of a16z for Uniswap to appear decentralized in terms of governance, while tight control rests with a16z. a16z could benefit from such a strategy in two possible ways. First, being the largest VC in Uniswap, the appearance of decentralization is likely to increase the value of UNI tokens if participants appreciate decentralization, which is consistent with our earlier result on the negative correlation of self-delegated votes and votes delegated from others. Second, a16z could also benefit from such window-dressing by gaining even tighter control.

[20] For economic interpretation, these coefficients must be converted back to their original scale. For example, the coefficient on a16z Affiliate (12.779 in specification 6 of Table 4) indicates that being an a16z affiliate is associated with an increase in votes delegated from others by a factor of e^12.779 - 1 = 354,689%. Converting this percentage increase to a multiplicative factor, we obtain 3,548 times more delegated votes compared to non-affiliates.

UNI holders other than a16z that appreciate decentralization would be more are likely to delegate to a16z affiliates than to a16z itself. Therefore, a16z can effectively through their affiliate network acquire more delegate votes than it normally would, increasing its benefits of control.

Austgen et al. (2023) argue that delegation enhances decentralization when voting power is delegated from one large, inactive token holder to multiple delegates who can vote independently, creating separate voting blocs. In our case, however, a16z uses its proprietary voting power by actively participating in governance, and its delegates vote in line with the venture capital firm. On the eleven events when Andreessen cast a vote at the critical on-chain vote level, only once did an affiliate vote in the opposing direction. a16z affiliates are also proposers. At the critical on-chain vote level, 36.36% of proposals are put forward by a16z affiliates. While this might suggest coordinated voting and proposing behaviour among a16z affiliates, other, non-affiliated Top5 voters display a similar level of voting alignment. Nevertheless, the combination of a16z's substantial direct voting power and their network of affiliates demonstrates their significant influence within the governance of Uniswap.

Our base specification also includes controls for market conditions. While most controls are statistically significant, they do not affect our main conclusions. Furthermore, none of the dummy variables on governance period exhibit statistical significance across specifications. This suggests that delegation decisions are not driven by individual voting events. Instead, they appear to be long-term decisions, akin to the appointment of independent board members in companies, who are designated to act as delegated monitors on behalf of shareholders.

Table 5 analyzes the influence of proposer-related characteristics on attracting voting power from others. We first observe that our previous findings on delegate characteristics remain robust in both magnitude and statistical significance after controlling for various proposal-related variables. Then, we find support for our third hypothesis. The variable, Was proposer, enters with a positive and statistically significant coefficient at the 1% level in the first specification. Economically, this implies that having previously put forward a proposal is associated with receiving approximately 59 times more votes than never having proposed, suggesting that delegators prefer to delegate voting power to those who not only vote but also actively engage in governance by submitting proposals. Notably, having previously proposed an on-chain vote is associated with receiving approximately 101 times more votes (specification 2), nearly double the effect size of general proposal activity across governance stages. In contrast, proposing at earlier governance stages does not yield statistically significant effects. This suggests that those who are able to propose at the final and most decisive stage of the governance process are significantly more attractive as delegates or more visible to delegators.

Proposing for an on-chain vote is likely correlated with prior success at earlier stages of the governance process. To examine whether proposal success is driving our results, we introduce a dummy variable in specification 3 that captures success at various stages. We find that the dummy variable enters with a positive and statistically significant coefficient at the 10% level in the specification. Economically, wallets associated with proposals that ultimately gain voter approval receive approximately 31 times more votes than others, suggesting that delegators consider proposer success, potentially as an indicator of the delegate's superior reputation.

Furthermore, in specification 4, we test whether having previously successfully proposed an on-chain vote is associated with receiving more votes compared to others. We find some evidence in support of this: the coefficient is positive but just falls short of statistical significance at the 10% level. Nonetheless, successfully proposing an on-chain vote is linked to receiving approximately 626 times more votes, whereas success at earlier governance stages does not yield a statistically significant effect. This finding reinforces the idea that delegators place the highest value on proposers who succeed at the final and most consequential stage of the decision-making process. In addition, the proposal count variables remain statistically insignificant in the remaining specifications of Table 5, indicating that it is proposal success—rather than mere experience—that matters to delegators.

Taken together, our results suggest that UNI token holders (and perhaps users) appreciate decentralization in governance processes and that a16z pretends to cater to that preference while holding tight control. In addition, voting power tends to concentrate around successful proposers, either due to reputation mechanisms or the inertia of vote delegations around widely supported proposals.[21]

## 5. Conclusion

Using Uniswap as a laboratory, we provide evidence consistent with three main drivers of vote delegations in DAOs: The token holders' desire for actual decentralization, reputation in actively and successfully making efforts to propel the DAO forward, and window dressing efforts regarding decentralization.

These findings highlight the tension between the promise and reality of decentralized governance, also when delegation of voting power is possible. While DAOs aim to distribute decision-making power, our analysis reveals that voting power can remain effectively concentrated amongst early-stage investors through a combination of holding tokens complemented by a strategic delegation network of aligned affiliates that cast their votes in the same direction.

For UNI token holders, our results reveal that their delegated voting power may inadvertently reinforce a16z's control even when delegating to an at first apparently independent delegate. Token holders who prioritize decentralization should carefully investigate delegate affiliations and consider active participation in governance rather than delegation.

For regulators and policymakers, our findings highlight a potential gap between the promise and reality of decentralized governance. The concentration of voting power through strategically aligned delegation networks raises questions about whether additional oversight of governance structures in DeFi platforms is warranted. Interestingly, such oversight could be implemented at the smart contract level as "embedded regulation" at relatively low cost (Zetsche et al., 2020).

[21] There may be herding in vote delegations. Such herding is consistent with reputation effects as long as "focal delegators" make their delegations based on quality delegates' decisions. Such herding dynamics could further enforce the influence of entities like a16z, as their support for a proposal can trigger cascading alignment, reinforcing the appearance of consensus while maintaining their effective control.

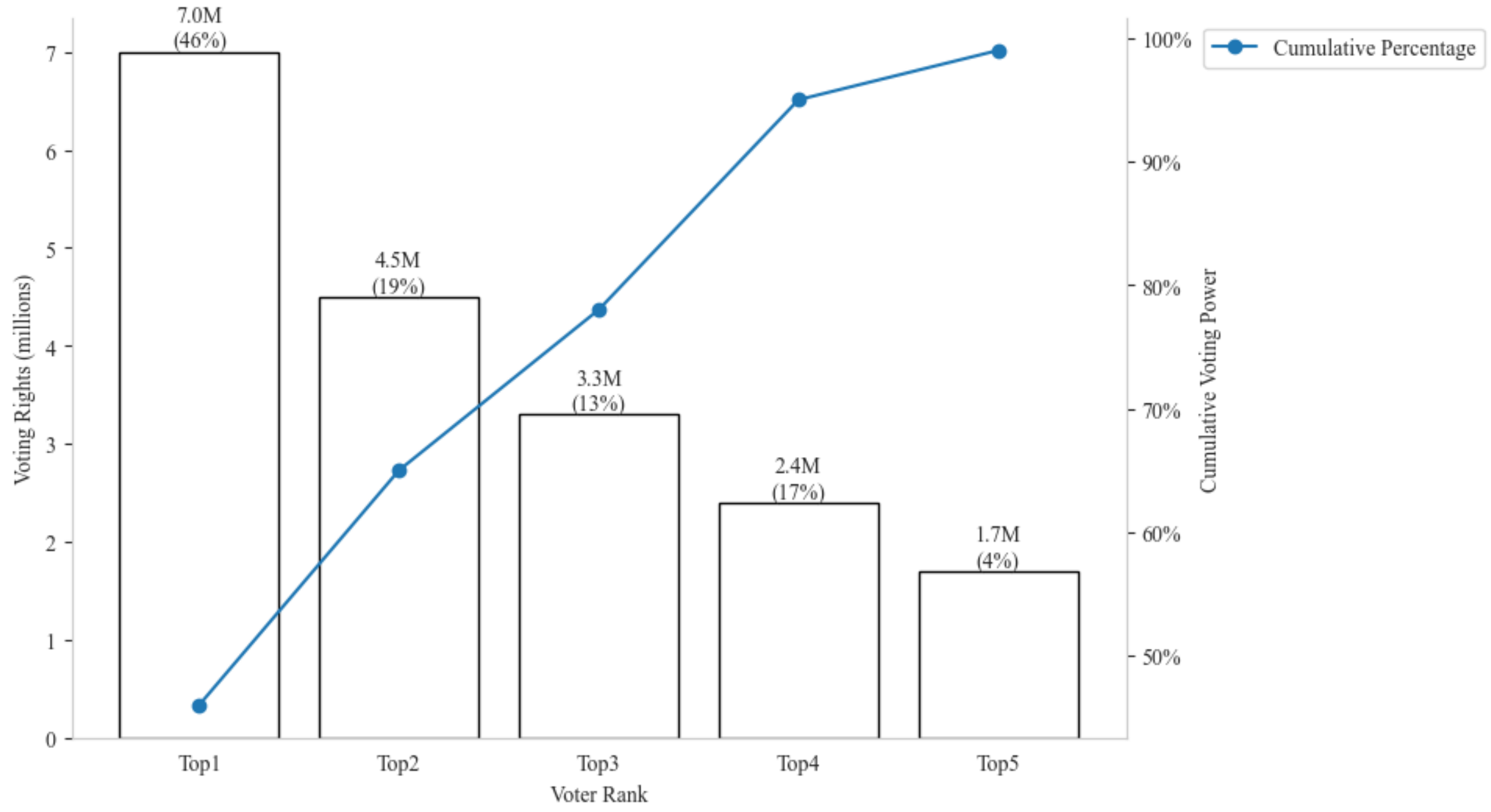

**Figure 1. Voting power.** This graph shows voting power in millions and as a percentage of votes cast for the "Top5" group (n=84) and overlays the cumulative voting power as a percentage of total votes cast. Both figures are averaged across 48 voting events. The "Top5" group consists of all unique delegates who ranked among the five largest voters in at least one voting event during our sample period between 18th of September 2020 to 15th of April 2022.

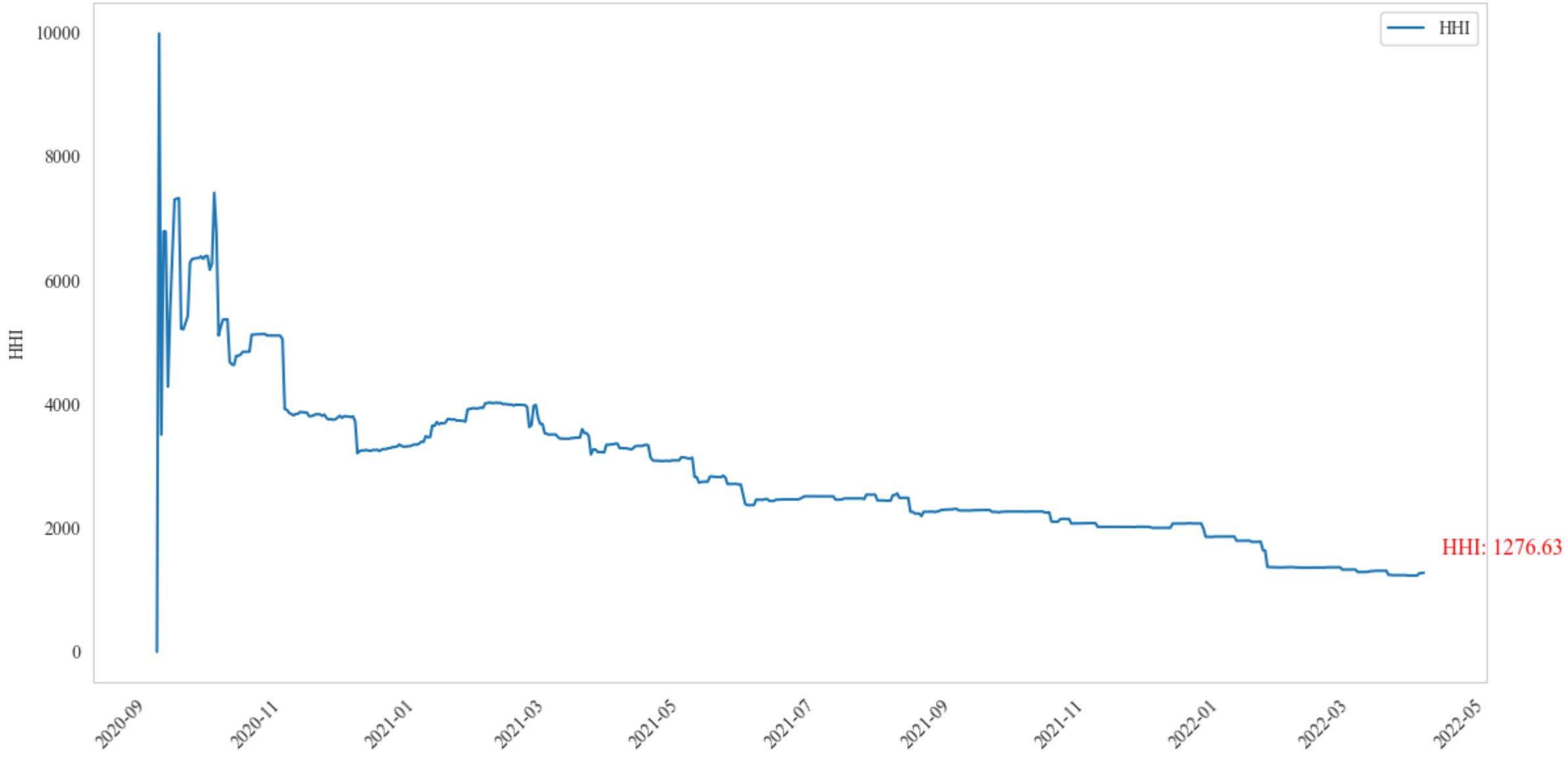

**Figure 2**. **HHI**. This figure shows Herfindahl-Hirschman Index (HHI) to show the concentration of voting power in Uniswap Governance ecosystem from 18th of September 2020 to 15th of April 2022. The index is calculated daily using delegate-level data on total votes delegated to them, where each delegate's market share is the proportion of total votes they control. a16z and its delegates, are grouped together, while smaller delegates are treated individually in the calculation.

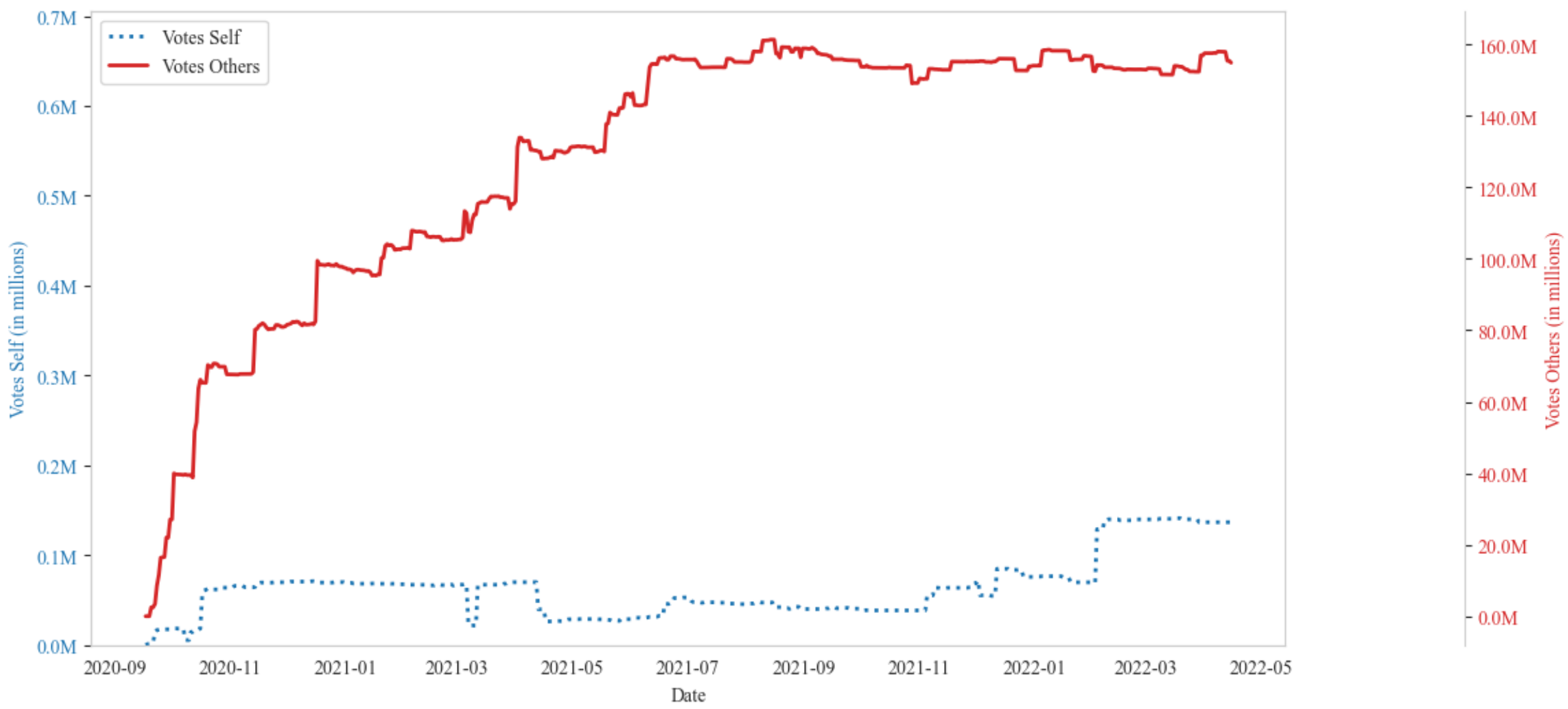

**Figure 3. Vote delegation to oneself vs. from others**. This chart shows two time series for the period from 18th of September 2020 to 15th of April 2022: Votes delegated to oneself, and votes delegated from others for the group of "Top5" wallets. The "Top5" group (n=84) comprises all unique delegates who ranked among the five largest voters in at least one voting event during our sample period.

**Table 1**. **Governance process stages**. This table lists the stages of the Uniswap Governance process, with corresponding voting share requirement to put forward a new proposal, the type of majority, required votes, and respective minimum duration. (Note: On 21.12.2022 the community voted to simplify the governance process as described here: https://gov.uniswap.org/t/community-governance-process-update-jan-2023/19976)

| **Process stage** | **UNI required to propose** | **Majority** | **Minimum Required votes** | **Minimum duration (days)** |
|---|---|---|---|---|
| Temperature Check | n/a | Simple majority | 25,000 affirmative votes | 2 |
| Consensus Check | n/a | Simple majority | 50,000 affirmative votes | 5 |
| On-chain Vote | 2,500,000 | Simple majority | 40,000,000 affirmative votes | 7 |

**Table 2**. **Variable description**.

| Variables | Sources | Descriptions |
|---|---|---|
| *Outcome variable* | | |
| Votes Delegated from Others | Dune / Ethereum Blockchain | Natural logarithm of a delegate's voting power plus one resulting from delegation to them from others on a given date |
| *Delegate characteristics variables* | | |
| Votes Delegated to Oneself | Dune / Ethereum Blockchain | Natural logarithm of a delegate's voting power plus one resulting from owning UNI tokens on a given date |
| Nickname | Etherscan | A dummy variable that is set to 1 if the delegate's Ethereum wallet address has a nickname and zero otherwise |
| a16z Affiliate | a16z website | A dummy variable that is set to 1 if the delegate qualifies as an a16z Affiliate (as listed on https://a16z.com/open-sourcing-our-token-delegate-program) and zero otherwise |
| *Market-related variables* | | |
| UNI Token Market Capitalization | CoinGecko | Natural logarithm of the total market capitalization of UNI on a given date in USD plus one |
| UNI Total Trading Volume | CoinGecko | Natural logarithm of the total traded volume across exchanges in the UNI token in USD plus one |
| Fear and Greed Index | Alternative.me | An index from zero (strong fear) to 100 (strong greed) indicating market sentiment and sourced from the alternative.me API |
| ETH Price | CoinGecko | Natural logarithm of the ETH price on a given date in USD plus one |
| Average Transaction Fee | Etherscan | Natural logarithm of the average gas fee in USD to be paid for transaction execution on the Ethereum blockchain plus one |
| *Voting -period-related variables* | | |
| Temperature Check Proposal Period | Messari | A dummy variable that is set to 1 if on a date a temperature check period is active and zero otherwise |
| Consensus Check Proposal Period | Messari | A dummy variable that is set to 1 if on a date a consensus check period is active and zero otherwise |
| On-chain Vote Proposal Period | Messari | A dummy variable that is set to 1 if on a date a on-chain vote period is active and zero otherwise |
| Any Proposal Period | Messari | A dummy variable that is set to 1 if on a date any of the above is active and zero otherwise |
| | | |
| *Proposer-related variables* | | |
| Was Proposer | Messari | A dummy variable that is set to 1 from the date when the proposer raised their first proposal and zero before that date |
| Was Proposer for Temperature Check | Messari | A dummy variable that is set to 1 from the date when the proposer raised their first temperature check proposal and zero before that date |
| Was Proposer for Consensus Check | Messari | A dummy variable that is set to 1 from the date when the proposer raised their first consensus check proposal and zero before that date |
| Was Proposer for On-chain Vote | Messari | A dummy variable that is set to 1 from the date when the proposer raised their first on-chain vote proposal and zero before that date |
| Was successful Proposer | Messari | A dummy variable that is set to 1 if a proposal of a proposer was successful and zero otherwise |
| Was successful Proposer for Temperature Check | Messari | A dummy variable that is set to 1 if a temperature check proposal of a proposer was successful and zero otherwise |
| Was successful Proposer for Consensus Check | Messari | A dummy variable that is set to 1 if a consensus check proposal of a proposer was successful and zero otherwise |
| Was successful proposer for On-chain Vote | Messari | A dummy variable that is set to 1 if a on-chain vote proposal of a proposer was successful and zero otherwise |
| | | |
| Temperature Check Proposal Count | Messari | The count of temperature check-level proposals a wallet has raised up until a given date |
| Consensus Check Proposal Count | Messari | The count of consensus check-level proposals a wallet has raised up until a given date |
| On-chain Vote Proposal Count | Messari | The count of on-chain vote-level proposals a wallet has raised up until a given date |
| Total Proposal Count | Messari | The sum of temperature check-level, consensus check-level and on-chain vote-level proposals a wallet has raised up until a given date |

**Table 3**. **Descriptive statistics**. This table reports descriptive statistics for the variables used in the study. All variables are described in Table 2. For variables with transformation, statistics marked with an asterisk (*) represent values calculated directly from the original raw data, not back-transformed from the log1p values.

| Variable | Mean | SD | Min | P50 | Max | Count |
|---|---|---|---|---|---|---|
| Votes Delegated from Others (Log1p) | 4.666 | 6.751 | 0.000 | 0.000 | 16.578 | 47,808 |
| | 1,371,854.090* | 2,885,099.54* | 0.000* | 0.000* | 15,845,481.7* | 47,808 |
| Votes Delegated to Oneself (Log1p) | 1.478 | 2.813 | 0.000 | 0.000 | 11.003 | 47,808 |
| | 443.670* | 2,676.810* | 0.000* | 0.000* | 60,081.000* | 47,808 |
| Nickname | 0.506 | 0.500 | 0.000 | 1.000 | 1.000 | 47,808 |
| a16z Affiliate | 0.181 | 0.385 | 0.000 | 0.000 | 1.000 | 47,808 |
| UNI Token Market Capitalization (Log1p) | 22.404 | 1.153 | 19.209 | 22.795 | 23.838 | 47,808 |
| | 8,375,665,113.54* | 5,776,527,375.87* | 219,913,212.5* | 7,942,139,164* | 22,531,045,773* | 47,808 |
| UNI Total Trading Volume (Log1p) | 19.865 | 0.778 | 18.424 | 19.743 | 22.462 | 47,808 |
| | 588,377,963.58* | 570,153,969.83* | 10,035,6151.1* | 375,200,043.9* | 5,688,408,417* | 47,808 |
| Fear and Greed Index | 0.540 | 0.258 | 0.100 | 0.530 | 0.950 | 47,808 |
| ETH Price (Log1p) | 7.562 | 0.761 | 5.775 | 7.838 | 8.480 | 47,808 |
| | 2,390.378* | 1,238.060* | 321.077* | 2,535.112* | 4,815.005* | 47,808 |
| Average Transaction Fee (Log1p) | 2.583 | 0.922 | 0.647 | 2.792 | 4.244 | 47,808 |
| | 17.937* | 14.511* | 0.910* | 15.320* | 68.720* | 47,808 |
| Temperature Check Proposal Period | 0.057 | 0.232 | 0.000 | 0.000 | 1.000 | 47,808 |
| Consensus Check Proposal Period | 0.083 | 0.276 | 0.000 | 0.000 | 1.000 | 47,808 |
| On-chain Vote Proposal Period | 0.083 | 0.276 | 0.000 | 0.000 | 1.000 | 47,808 |
| Any Proposal Period | 0.167 | 0.373 | 0.000 | 0.000 | 1.000 | 47,808 |
| Was Proposer | 0.088 | 0.284 | 0.000 | 0.000 | 1.000 | 47,808 |
| Was Proposer for Temperature Check | 0.055 | 0.227 | 0.000 | 0.000 | 1.000 | 47,808 |
| Was Proposer for Consensus Check | 0.035 | 0.183 | 0.000 | 0.000 | 1.000 | 47,808 |
| Was Proposer for On-chain Vote | 0.036 | 0.187 | 0.000 | 0.000 | 1.000 | 47,808 |
| Was Successful Proposer | 0.055 | 0.229 | 0.000 | 0.000 | 1.000 | 47,808 |
| Was Successful Proposer for Temperature Check | 0.034 | 0.181 | 0.000 | 0.000 | 1.000 | 47,808 |
| Was Successful Proposer for Consensus Check | 0.023 | 0.149 | 0.000 | 0.000 | 1.000 | 47,808 |
| Was Successful Proposer for On-chain Vote | 0.024 | 0.154 | 0.000 | 0.000 | 1.000 | 47,808 |
| Temperature Check Proposal Count | 0.157 | 0.363 | 0.000 | 0.000 | 1.000 | 47,808 |
| Consensus Check Proposal Count | 0.108 | 0.311 | 0.000 | 0.000 | 1.000 | 47,808 |
| On-chain Vote Proposal Count | 0.072 | 0.259 | 0.000 | 0.000 | 1.000 | 47,808 |
| Total Proposal Count | 0.337 | 0.749 | 0.000 | 0.000 | 3.000 | 47,808 |

**Table 4**. **Delegated votes from others**. This table reports coefficient estimates from Tobit regressions of the natural logarithm Votes delegated from others plus one. All variables are described in Table 2. Sample sizes vary based on filters applied to exclude proposer records to avoid bias. Numbers in square brackets are standard errors clustered at the delegate level. *, **, and *** indicate significance at the 10%, 5%, and 1% levels, respectively.

| **Dependent variable** | **Spec1** | **Spec2** | **Spec3** | **Spec4** | **Spec5** | **Spec6** |
|---|---|---|---|---|---|---|
| Votes Delegated from Others | [b/se] | [b/se] | [b/se] | [b/se] | [b/se] | [b/se] |
| Votes Delegated to Oneself | -1.584*** | -1.625*** | -1.625*** | -1.625*** | -1.625*** | -1.625*** |
| | [0.527] | [0.532] | [0.532] | [0.532] | [0.532] | [0.532] |
| Nickname | 4.537 | 4.557 | 4.558 | 4.56 | 4.56 | 4.558 |
| | [3.344] | [3.381] | [3.381] | [3.381] | [3.381] | [3.381] |
| a16z Affiliate | 12.502*** | 12.783*** | 12.780*** | 12.776*** | 12.776*** | 12.779*** |
| | [2.765] | [2.793] | [2.793] | [2.792] | [2.792] | [2.793] |
| Average Transaction Fee | -1.258*** | -1.186*** | -1.181*** | -1.193*** | -1.170*** | -1.190*** |
| | [0.283] | [0.294] | [0.293] | [0.295] | [0.290] | [0.296] |
| ETH Price | 10.122*** | 10.032*** | 10.109*** | 10.171*** | 10.166*** | 10.107*** |
| | [1.557] | [1.560] | [1.552] | [1.548] | [1.535] | [1.524] |
| UNI Token Market Capitalization | -1.493** | -1.496** | -1.558** | -1.617** | -1.634** | -1.551** |
| | [0.692] | [0.671] | [0.666] | [0.670] | [0.663] | [0.653] |
| UNI Total Trading Volume | -1.388*** | -1.380*** | -1.372*** | -1.386*** | -1.382*** | -1.379*** |
| | [0.200] | [0.209] | [0.209] | [0.208] | [0.207] | [0.210] |
| Fear and Greed Index | -2.855** | -2.919** | -2.909** | -2.850** | -2.844** | -2.893** |
| | [1.377] | [1.393] | [1.393] | [1.410] | [1.410] | [1.407] |
| Temperature Check Proposal Period | | 0.126 | | | 0.285 | |
| | | [0.572] | | | [0.564] | |
| Consensus Check Proposal Period | | | -0.226 | | -0.134 | |
| | | | [0.321] | | [0.298] | |
| On-chain Vote Proposal Period | | | | -0.674 | -0.727 | |
| | | | | [0.572] | [0.528] | |
| Any Proposal Period | | | | | | -0.133 |
| | | | | | | [0.375] |
| Week FE | YES | YES | YES | YES | YES | YES |
| Observations | 47808 | 46840 | 46840 | 46840 | 46840 | 46840 |
| Pseudo R-squared | 0.098 | 0.099 | 0.099 | 0.099 | 0.099 | 0.099 |

**Table 5**. **Delegated votes from others and proposer-related variables**. This table reports coefficient estimates from Tobit regressions of the natural logarithm of Votes delegated from others plus one. All variables are described in Table 2. Numbers in square brackets are standard errors clustered at the delegate level. *, **, and *** indicate significance at the 10%, 5%, and 1% levels, respectively.

| **Dependent variable**<br>Votes Delegated from Others | **Spec1**<br>[b/se] | **Spec2**<br>[b/se] | **Spec3**<br>[b/se] | **Spec4**<br>[b/se] | **Spec5**<br>[b/se] | **Spec6**<br>[b/se] |
|---|---|---|---|---|---|---|
| Votes Delegated to Oneself | -1.591*** | -1.600*** | -1.575*** | -1.597*** | -1.600*** | -1.637*** |
| | [0.501] | [0.508] | [0.504] | [0.500] | [0.530] | [0.522] |
| Nickname | 4.944 | 5.079 | 4.419 | 4.999 | 4.41 | 4.415 |
| | [3.244] | [3.241] | [3.269] | [3.280] | [3.360] | [3.318] |
| a16z Affiliate | 10.916*** | 10.480*** | 11.640*** | 11.521*** | 12.893*** | 12.313*** |
| | [2.839] | [2.914] | [2.868] | [2.851] | [2.869] | [2.849] |
| Average Transaction Fee | -1.207*** | -1.215*** | -1.236*** | -1.214*** | -1.255*** | -1.253*** |
| | [0.286] | [0.286] | [0.281] | [0.286] | [0.281] | [0.279] |
| ETH Price | 8.892*** | 9.053*** | 9.067*** | 9.433*** | 10.123*** | 10.103*** |
| | [1.515] | [1.533] | [1.482] | [1.451] | [1.557] | [1.555] |
| UNI Token Market Capitalization | -1.015 | -1.059 | -1.111* | -1.259** | -1.496** | -1.510** |
| | [0.654] | [0.654] | [0.654] | [0.636] | [0.693] | [0.694] |
| UNI Total Trading Volume | -1.281*** | -1.318*** | -1.306*** | -1.356*** | -1.389*** | -1.396*** |
| | [0.212] | [0.208] | [0.209] | [0.200] | [0.200] | [0.200] |
| Fear and Greed Index | -2.485* | -2.629* | -2.567* | -2.897** | -2.871** | -2.927** |
| | [1.364] | [1.346] | [1.362] | [1.344] | [1.373] | [1.361] |
| Was Proposer | 8.670*** | | | | | |
| | [3.188] | | | | | |
| Was Proposer for Temperature Check | | 5.142 | | | | |
| | | [5.482] | | | | |
| Was Proposer for Consensus Check | | 0.298 | | | | |
| | | [5.767] | | | | |
| Was Proposer for On-chain Vote | | 9.207* | | | | |
| | | [5.039] | | | | |
| Was Successful Proposer | | | 8.020* | | | |
| | | | [4.207] | | | |
| Was Successful Proposer for Temperature Check | | | | 5.911 | | |
| | | | | [6.228] | | |
| Was Successful Proposer for Consensus Check | | | | -7.633 | | |
| | | | | [6.837] | | |
| Was Successful Proposer for On-chain Vote | | | | 11.043 | | |
| | | | | [6.755] | | |
| Total Proposal Count | | | | | -0.672 | |
| | | | | | [1.495] | |
| Temperature Check Proposal Count | | | | | | 0.81 |
| | | | | | | [3.889] |
| Consensus Check Proposal Count | | | | | | -5.422 |
| | | | | | | [5.420] |
| On-chain Vote Proposal Count | | | | | | 4.194 |
| | | | | | | [5.455] |
| Week FE | YES | YES | YES | YES | YES | YES |
| Observations | 47808 | 47808 | 47808 | 47808 | 47808 | 47808 |
| Pseudo R-squared | 0.108 | 0.106 | 0.104 | 0.104 | 0.099 | 0.101 |

## Appendix A: Details on Delegation Mechanism

### Delegation demonstration

We demonstrate that delegated voting power within the Uniswap DAO does not change when UNI tokens are transferred from one wallet to another. To set up our demonstration, we create three Ethereum wallets. Uni-Wallet01, Uni-Wallet02 and Uni-Wallet03. We fund Uni-Wallet01 and Uni-Wallet02 with a small amount of ETH to be able to pay gas fees associated with the functions we need to perform. We make use of the *transfer*, *swap*, and *delegate* functions.

**Table A1. Delegation demonstration wallet details**.

| Wallet Name | Public Key | Function |
|---|---|---|
| Uni-Wallet01 | 0xe245eA7953a0F09674531b8a92b7d7e3eeD0eD97 | Delegator |
| Uni-Wallet02 | 0xfe9fE83c9ae73DfFD2525F60e534c6d51074ad8a | Delegate |
| Uni-Wallet03 | 0xFa491ab7D82f4C6D19791A98e7afa816BE164792 | New Owner |

In step 1 we confirm that Uni-Wallet01 and Uni-Wallet02 are funded with small amounts of ETH by checking the corresponding transaction hash records on etherscan.io.[22] In step 2 we purchase UNI tokens. We first connect Uni-Wallet01 with the Uniswap decentralized exchange website and purchase one UNI token. One UNI token represents one voting right. Next, we confirm if the one UNI token has been credited to Uni-Wallet01 by checking the corresponding transaction hash record on etherscan.io. Thereafter, we delegate. Uni-Wallet01 can either delegate its voting right to itself or to another Ethereum wallet. We connect to Uni-Wallet01 to the Governance module of the Uniswap website and chose to delegate the one voting right to Uni-Wallet02. We confirm that the delegation of the one voting right is successful by checking the logs of the corresponding record on etherscan.io. Now we execute the token transfer. We transfer the one UNI token from Uni-Wallet01 to Uni-Wallet03. We confirm that the transfer of the one UNI token is successful by checking the corresponding record on etherscan.io.

In the final step, we vote on a proposal with Uni-Wallet02 to confirm the voting right remains with Uni-Wallet02, even after the transfer of the UNI token from Uni-Wallet01 to Uni-Wallet03. Once again, the transaction hash of this vote can be verified using etherscan.io. We then successfully cast a vote on an active proposal using Uni-Wallet02, confirming that the voting power remains with the wallet despite of the UNI token having been transferred to Uni-Wallet03 already.

[22] The public keys of these wallets are listed in Table XX. Anyone can inspect these wallets and respective transaction hashes on any Ethereum explorer like etherscan.io

**Technical smart contract analysis**

We now conduct an analysis of the UNI token contract highlighting the technical details behind phenomenon.

When analyzing how voting power moves during transfers, we examine what happens when Uni-Wallet01 transfers tokens to Uni-Wallet03. On line 2, we start with the condition `if (srcRep != dstRep && amount > 0)`. The issue arises because delegates [Uni-Wallet03] will be `address(0)` if Uni-Wallet03 has never delegated before. The critical behavior occurs on line 11 with the condition `if (dstRep != address(0))`. Since Uni-Wallet03's delegate is the zero address, this condition fails, meaning lines 13-16 never execute. However, if voting power was previously delegated to Uni-Wallet02, the source delegate's voting power adjustment still occurs in lines 3-9, where the votes would typically be subtracted from Uni-Wallet02's total. This creates a situation where voting power should be removed from Uni-Wallet02 but it is not, allowing Uni-Wallet02 to retain the voting power until Uni-Wallet03 explicitly delegates their voting power (either to themselves or another address).[23] The solidity function in question:

```
Line Code
1    function _moveDelegates(address srcRep, address dstRep, uint96 amount) internal {
2        if (srcRep != dstRep && amount > 0) {
3            if (srcRep != address(0)) {
4                // Reduce old delegate's voting power
5                uint32 srcRepNum = numCheckpoints[srcRep];
6                uint96 srcRepOld = srcRepNum > 0 ? checkpoints[srcRep][srcRepNum - 1].votes : 0;
7                uint96 srcRepNew = sub96(srcRepOld, amount, "Uni::_moveVotes: vote amount underflows");
8                _writeCheckpoint(srcRep, srcRepNum, srcRepOld, srcRepNew);
9            }
10
11           if (dstRep != address(0)) {
12               // Increase new delegate's voting power
13               uint32 dstRepNum = numCheckpoints[dstRep];
14               uint96 dstRepOld = dstRepNum > 0 ? checkpoints[dstRep][dstRepNum - 1].votes : 0;
15               uint96 dstRepNew = add96(dstRepOld, amount, "Uni::_moveVotes: vote amount overflows");
16               _writeCheckpoint(dstRep, dstRepNum, dstRepOld, dstRepNew);
17           }
18       }
19   }
```

[23] If Uni-Wallet01transfers tokens to Uni-Wallet03 but Uni-Wallet03 already contains UNI tokens and has these already delegated, the additional tokens are automatically delegated in the same way as the existing ones that Uni-Wallet03 already owns.

## Appendix B: Additional Figure

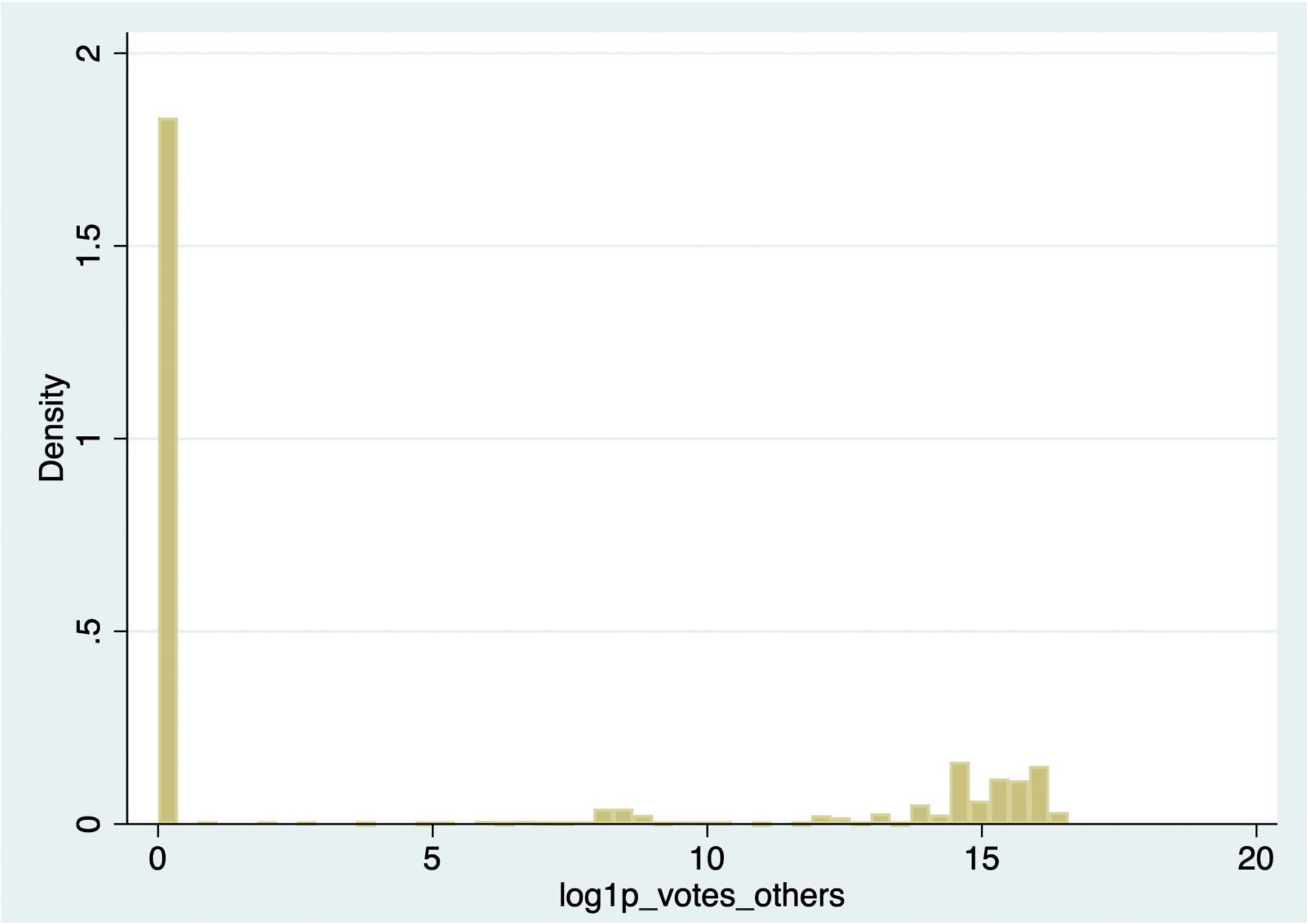

**Figure A1. Density plot.** This figure shows a density plot of the log-transformed outcome variable Votes Delegated from Others for the group of "Top5" wallets. The "Top5" group (n=84) comprises all unique delegates who ranked among the five largest voters in at least one voting event during our sample period of 18th of September 2020 to 15th of April 2022.